\documentstyle[preprint,aps,prb,epsf]{revtex}

\newcommand\ltappr{{{\lower4pt\hbox{$<$} } \atop \widetilde{ \ \ \ }}}

\begin{document}

\title{Break up of the heavy electron at a quantum critical point}

\author{J. Custers$^{\ast}$, P. Gegenwart$^{\ast}$, H. Wilhelm$^{\ast}$, K. Neumaier${\dag}$,
Y. Tokiwa$^{\ast}$, O. Trovarelli$^{\ast}$, C. Geibel$^{\ast}$, F. Steglich$^{\ast}$, C. P{\'e}pin${\ddag}$ \& P.
Coleman${\S}$}

\address{$^{\ast}$Max-Planck-Institute for the Chemical Physics of Solids, D-01187 Dresden, Germany}
\address{${\dag}$Walther-Meissner-Institute for Low Temperature Research of the Bavarian Academy of Sciences, D-85748 Garching, Germany}
%\address{${\ddag}$Graduate School of Science, Osaka University, Toyonaka, Osaka 560-0043, Japan}
\address{${\ddag}$SPhT, L'Orme des Merisiers, CEA-Saclay, 91190 Gif-sur-Yvette, France}
\address{${\S}$CMT, Department of Physics and Astronomy, Rutgers University, Piscataway, NJ 08854-8019, USA}

\date{\today}

\maketitle

{\bf
%%"The point at absolute zero ... order abd disorder^1-4. Previously it has been
%%determined that ... (explain clearly what was done in ref. 18), but that work left
%%unresolved the important issue(s) of ... (explain). Here we report ... These results
%%allow us to (specify how this resolves the issues left unresolved by ref. 18)."
%
The point at absolute zero where matter becomes unstable to new forms
of order is called a quantum critical point (QCP).  The quantum
fluctuations between order and disorder\cite{PRB1976Hertz,PRB1993Millis,PR1994Continentino,NAT2001Si,qcp} that develop at this
point induce profound transformations in the finite temperature electronic
properties of the material.  Magnetic fields are ideal for tuning a
material as close as possible to a QCP, where the most intense effects
of criticality can be studied.  A previous study\cite{PRL2002Gegenwart} on theheavy-electron material $YbRh_2Si_2$ found that near a field-induced
quantum critical point electrons move ever more slowly and scatter off
one-another with ever increasing probability, as indicated by a
divergence to infinity of the electron effective mass and
cross-section.  These studies could not shed light on whether these
properties were an artifact of the applied field\cite{greg,hilbert}, or 
a more general feature of field-free QCPs.  Here we report that when 
Germanium-doped $YbRh_2Si_2$ is tuned away from a chemically induced quantum 
critical point by magnetic fields there is a universal behavior in the
temperature dependence of the specific heat and resistivity:  
the characteristic kinetic energy of electrons is directly
proportional to the strength of the applied field.  We infer that all
ballistic motion of electrons vanishes at a QCP, forming a new class of
conductor in which individual electrons decay into collective current
carrying motions of the electron fluid.
}

Recent work\cite{PRL2002Gegenwart} on the heavy electron material
YbRh$_{2}$Si$_{2}$~\cite{PRL2000Trovarelli} has demonstrated that
a magnetic field can be used to probe the heavy electron quantum
critical point. This material exhibits a small antiferromagnetic
(AFM) ordering temperature $T_{\rm N}=70$~mK (Fig.~1a) that is
driven to zero by a critical magnetic field $B_{\rm c}=0.66$ T (if
the field is applied parallel to the crystallographic $c$--axis,
perpendicular to the easy magnetic plane)\cite{PRL2002Gegenwart}.
%The various Fermi liquid parameters of the  material, such as the linear temperature coefficient of
% the specific heat $\gamma$ and the quadratic coefficient of the resistivity $A$ were found to
% diverge in the approach to the quantum critical point. This  dramatic result contradicts
%the standard spin density wave model of a magnetic quantum critical point. One of the great difficulties with
%these results is that they were carried out at a finite field QCP.
For $B>B_{\rm c}$, a field-induced Landau Fermi Liquid (LFL) state       %commentaar: . into , and subscript "c" in roman style%
characterized by $\Delta\rho=AT^2$ (where $\Delta \rho(T) =
\rho(T)-\rho_{o}$ is the temperature dependent part of the
electrical resistivity) is established below some cross-over
temperature $T_0(B)$ which grows linearly with field. The $A$
coefficient, being proportional to the quasiparticle-quasiparticle scattering cross
section was found to diverge as $A(B)\propto 1/(B-B_{\rm c})$ for  %commentaar: Sigma deleted by JC and subscript "c" in roman style%
$B\rightarrow B_{\rm c}$. Comparative studies of the resistivity
and the electronic specific heat $C_{\rm {el}}(T)=\gamma_{o}(B)T$
in the field ranges 0.5~T~--~4~T ($B\perp c$ with $B_{\rm c}=0.06$ T)  %commentaar: subscript "c" in roman style%
and 2~T~--~6~T ($B\parallel c$) revealed a field-independent ratio
$A/\gamma_0^2$ slightly smaller than the empirical Kadowaki-Woods
ratio \cite{SSCom1986Kadowaki} that holds for LFL systems. This
seemed to suggest a divergence of the effective quasiparticle (QP) mass as
$1/\left(B-B_{\rm c}\right)^{1/2}$ as $B\rightarrow B_{\rm c}$. In this             %commentaar: subscript "c" in roman style%
letter, we report the first-ever observation of the divergence of
the QP mass at a QCP, established very close to $B=0$.

\begin{figure}
  \begin{center}
  \epsfxsize=85mm \epsfbox{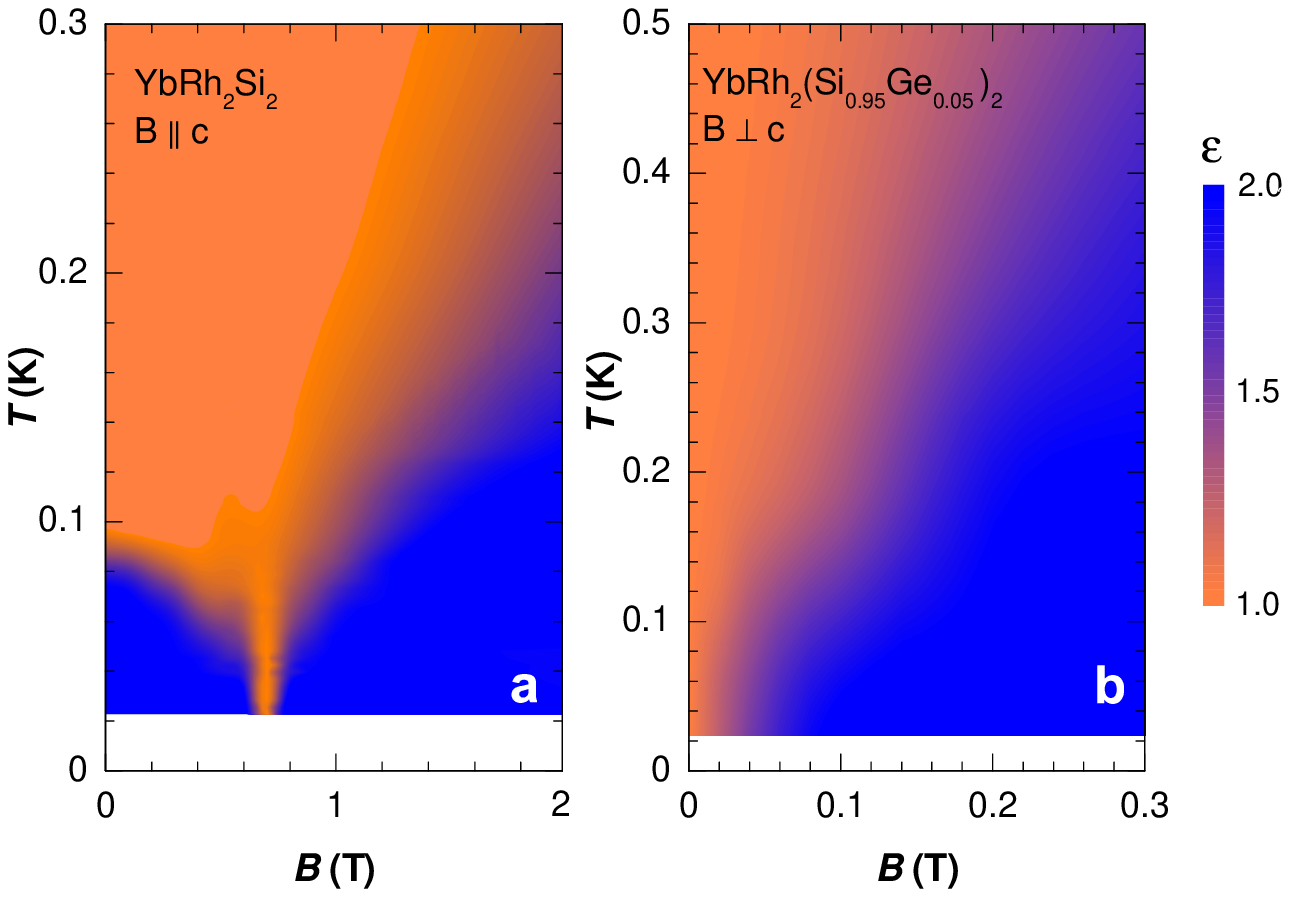}            %commentaar:Name of Figure changed!!!%
\end{center}
\vspace*{2mm} \caption{Evolution of $\varepsilon$, the exponent
in $\Delta \rho(T)=\left[\rho(T)-\rho_{o}\right]\propto
T^{\varepsilon}$, within the temperature--field phase diagram of
YbRh$_2$(Si$_{1-x}$Ge$_{x}$)${_2}$ single crystals. The
non--Fermi--liquid (NFL) behavior, $\varepsilon = 1$ (yellow
color), is found to occur at the lowest temperatures right at the
quantum critical point (QCP), $B=B_{\rm c}${ }$\big[$ {\bf a,}
$x=0$, $B_{\rm c} = 0.66$~T $\left(\parallel c\right)$, residual
resistivity $\rho_{o}=1\mu\Omega$cm; {\bf b,} $x=0.05$, $B_{\rm
c}=0.027$~T $\left(\perp c\right)$, $\rho_{o} = 5\mu\Omega$cm
$\big]$ and in a largely extended field range at higher
temperatures. For $B>B_{\rm c}$, a broad cross--over regime from
the NFL state to the field--induced heavy Landau Fermi--liquid
(LFL) state (at lower temperature) is stated. The LFL state is
characterized by $\Delta \rho(T) \propto T^{\varepsilon}$,
$\varepsilon =2$ (blue color). As shown in ({\bf a}) the
antiferromagnetically ordered phase of pure YbRh$_{2}$Si$_{2}$
below $T_{\rm N} = 70$~mK and $B_{\rm c}$ shows, owing to an
extremely small ordered moment, the outward appearance of a heavy
LFL state, too. Its phase boundary to the paramagnetic state is
manifested by a rapid change in $\varepsilon$ from 2 to 1.\\ The
low ordering temperature of pure YbRh$_{2}$Si$_{2}$ increases as
external pressure is applied~\protect\cite{PRL2000Trovarelli}.
The extrapolation of $T_{\rm N}(p) \rightarrow 0$ yields a
critical pressure $p_{\rm {c}}=-0.3(1)$~GPa, reflecting that a
small expansion of the unit cell volume, $V$, would tune $T_{\rm
N} \rightarrow 0$. This can be achieved by the substitution of Si
by the isoelectronic, but larger,
Ge~\protect\cite{JPCM2002Mederle}. Studies of the electrical
resistivity under pressure revealed a $T_{\rm N} \propto
(p+p_{\rm c})^{n}$ variation, with $n = 1.33$ for both compounds.
The $T_{\rm N}(p)$-dependences of the $x=0$ and $x=0.05$ crystals
can be matched if all $x=0.05$ data points are shifted by the
same amount $\Delta p = -0.17(2)$~GPa to lower
pressure~\protect\cite{JPCM2002Mederle}, yielding $T_{\rm N}=
20(5)$~mK. Using the bulk modulus $B_0=189$~GPa of
YbRh$_{2}$Si$_{2}$~\protect\cite{Abd-Elmeguid1999}, the small
pressure shift of $\Delta p = -0.17(2)$~GPa is equivalent to a
volume expansion of $\Delta V = 0.14(3)$~\AA$^{3}$. This
transforms into an {\it effective} Ge content $x_{\rm {eff}} =
0.019(6)$, if the value $\Delta V / V=7.65(78)$~\% for the
relative change of the unit-cell volume with Ge concentration in
the solid-solution YbRh$_{2}$(Si$_{1-x}$Ge$_{x}$)$_{2}$ is used,
with $V(x=0)=158.4(2)$~\AA$^3$ and $V(x=1)=166.07(54)$~\AA$^3$,
cf. Ref.~\protect\cite{Francois1985}, in agreement with microprobe
analysis~\protect\cite{APPB2003Gegenwart}.}
\label{fig:JCNatfiguur3}
\end{figure}

\begin{figure}
\begin{center} \epsfxsize=85mm \epsfbox{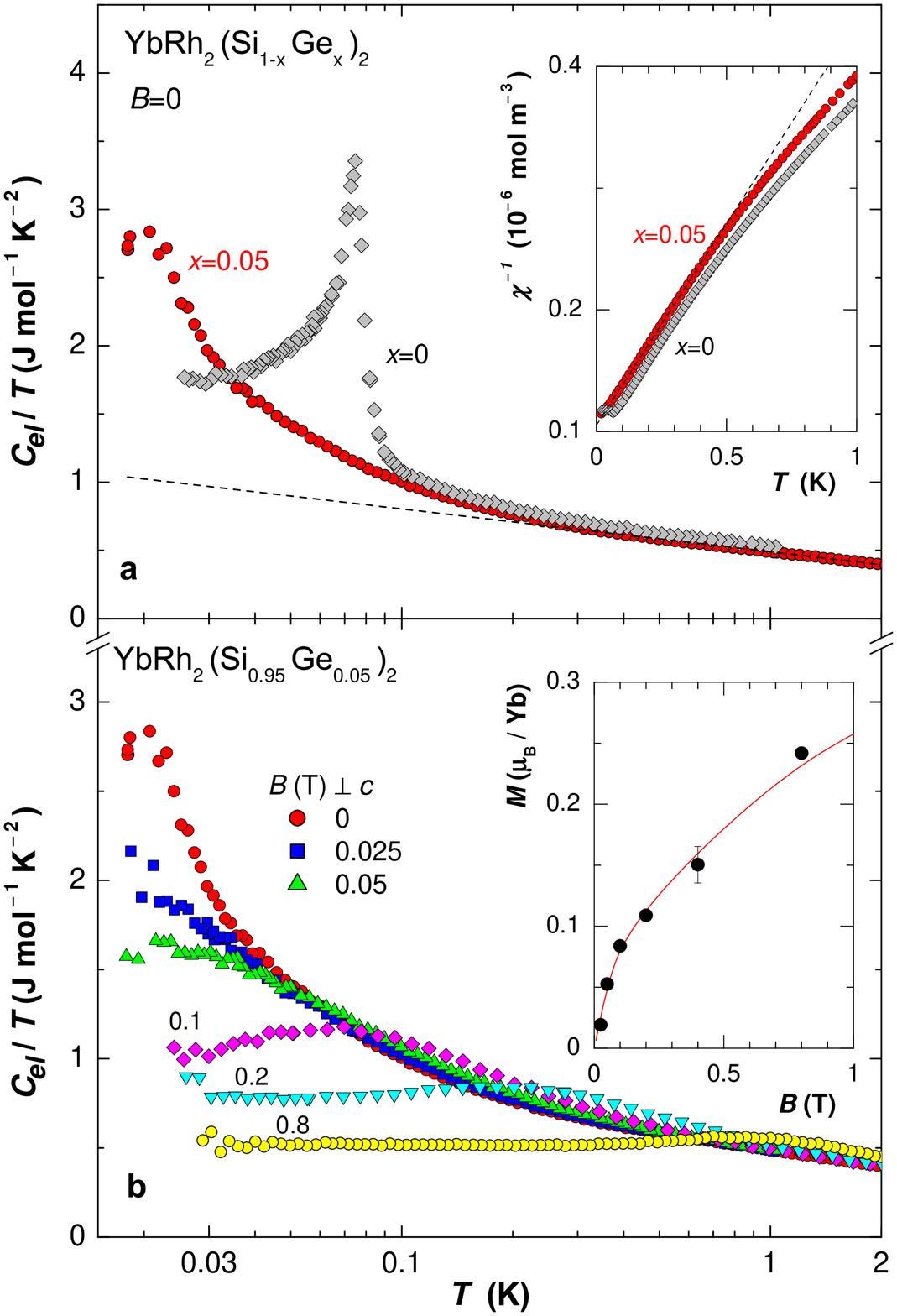}           %commentaar:Name of Figure changed!!!%
\end{center}
\vspace*{2mm} \caption{Low--temperature electronic specific heat
of YbRh$_{2}$(Si$_{1-x}$Ge$_{x}$)$_{2}$ single crystals as $C_{\rm
{el}}/T$ vs $T$ in semi--log plots at zero field and at low values
of the applied magnetic field $B$. Insets show low--$T${ }$B=0$
ac--susceptibility as $\chi^{-1}$ vs $T$ ({\bf a}) and
magnetization as $M$ vs $B$ ({\bf b}). $C_{\rm {el}}$ is obtained
by subtracting the nuclear quadrupolar contribution, $C_{\rm Q} =
\alpha_{\rm Q}/T^{2}$ (with $\alpha_{\rm Q} = 5.68 \times
10^{-6}$~JKmol$^{-1}$, calculated from recent M{\"o}ssbauer
results~\protect\cite{Abd-Elmeguid1999}) ({\bf a}) and, in
addition, the nuclear Zeeman contribution $C_{\rm
hf}=\alpha(B)/T^{2}$ ({\bf b}), from the raw data. Here,
$\alpha(B)$ has been deduced by plotting $CT^2$ vs $T^3$. The
magnetization, $M$ vs magnetic field $B$ (black points in the
inset), is calculated via ($B_{hf}-B)/A$, with $A$ the hyperfine
coupling constant for Yb in this compound and the hyperfine field
$B_{hf}=\sqrt{(\alpha(B)-\alpha_Q)/\alpha_{dip}}$; $\alpha_{dip}$
represents the strength of the nuclear magnetic dipolar
interaction and amounts to $7.58\times
10^{-8}$~JKmol$^{-1}$T$^{-2}$, Ref.~\protect\cite{PMS1977Carter}.
With the assumption of $A = 120$~T/$\mu_{\rm B}$, the data points
agree perfectly with the measured magnetization curve at 40~mK
(red line in the inset of ({\bf b}).\\
The $B=0$ results shown in ({\bf a}) reveal an upturn in $C_{\rm
el}(T)/T$ for paramagnetic YbRh$_{2}$(Si$_{1-x}$Ge$_{x}$)$_{2}$
($x=0$, $T_{\rm N}=70$~mK; $x=0.05$, $T_{\rm N}=20$~mK) below
$T=0.3$~K. In the same temperature range the susceptibility
$\chi(T)$ shows a Curie--Weiss law, $\chi^{-1} \propto
(T-\Theta)$ [inset of ({\bf a})]. For both samples very similar
values are found for the Weiss temperature, $\Theta \approx
-0.3$~K, as well as for the large effective moment, $\mu_{\rm
eff} \approx 1.4 \mu_{\rm B}/$Yb$^{3+}$. For
YbRh$_{2}$(Si$_{0.95}$Ge$_{0.05}$)$_{2}$, entropy is shifted from
low to higher temperatures when a magnetic field is applied ({\bf
b}). The cross--over temperature between the field--induced LFL
state ($C_{\rm el}(T)/T \approx const.$) and the NFL state at
higher temperature is depicted by the position of the broad hump
in $C_{\rm el}(T)/T$ which shifts upwards linearly with the
field, $B\leq 0.8$~T ($\perp c$).} \label{fig:specHeatacSuscep}
\end{figure}

By alloying YbRh$_{2}$Si$_{2}$ with Germanium, using a nominal
concentration $x=0.05$, we have been able to fine-tune, in a new
set of studies, the N{\'e}el temperature of this material and the
critical field far closer to zero, to a point where, for the
first time,  we may reliably probe the zero-field transition
using field-tuning. The phase diagram for a high--quality
YbRh$_{2}$(Si$_{0.95}$Ge$_{0.05}$)$_{2}$ single crystal is shown
in Fig.~\ref{fig:JCNatfiguur3}b. NFL behavior dominates over a
funnel-shaped region of the $T$--$B$ phase diagram down to the
lowest accessible temperature of $20$~mK. The critical field has
been suppressed to as low as $B_{\rm c}=0.027$ T ($B\perp c$). As       %commentaar: subscript "c" in roman style%
in the undoped material, there is a broad cross--over regime
between the NFL and field polarized LFL regime with a mean
cross--over temperature $T_{o}$ that is seen to rise linearly
with the field $B$. Very weak AFM order develops in the $x=0.05$         %commentaar: field "H" into field "B"%
sample below $T_{\rm N}=20$~mK, as evidenced by the extremely
weak anomaly in the electronic specific heat coefficient
(Fig.~\ref{fig:specHeatacSuscep}a).

Past experience\cite{greg,hilbert} suggested that a finite field
quantum critical point has properties which are qualitatively
different to a zero field transition, shedding doubt on the
reliability of these measurements as an indicator of the physics
of a quantum phase transition at zero field. However, the
zero--field properties of YbRh$_{2}$(Si$_{1-x}$Ge$_{x}$)$_{2}$         %commentaar: YbRh2(Si0.95Ge0.05)2 changed in YbRh2(Si1-xGex)2%
above $T \approx 70$~mK for the undoped ($x=0$) and doped             %commentaar: T>0.3K into T~70mK%
($x=0.05$) crystals are essentially identical
(Fig.~\ref{fig:specHeatacSuscep}a), suggesting that by
suppressing the critical field we are still probing the same
quantum critical point. In both compounds, the ac-susceptibility
follows a temperature dependence $\chi^{-1} \propto T^{\alpha}$
from $0.3$~K to $\leq T \leq 1.5$~K, with $\alpha =
0.75$~\cite{APPB2003Gegenwart}, and the coefficient of the
electronic specific heat, $C_{\rm {el}}(T)/T$,
exhibits\cite{PRL2000Trovarelli} a logarithmic divergence between
0.3~K and 10~K. However, in the low-$T$ paramagnetic regime,
i.\,e.\,, $T_{\rm N} < T \lesssim 0.3$~K, the ac-susceptibility
follows a Curie-Weiss law (inset of
Fig.~\ref{fig:specHeatacSuscep}a) with a Weiss temperature
$\Theta_{W}\approx 0.3$~K, and a surprisingly large effective
moment $\mu_{\rm {eff}} \approx 1.4 \mu_{\rm B}/$Yb$^{3+}$,              %commentaar: mu_B  into mu_B/Yb^3+%
indicating the emergence of coupled, unquenched spins at the
quantum critical point. The electronic specific heat coefficient,
$C_{\rm {el}}(T)/T$, exhibits a pronounced upturn below $0.3$~K
(Fig.~\ref{fig:specHeatacSuscep}a).

\begin{figure}
  \begin{center}
  \epsfxsize=85mm \epsfbox{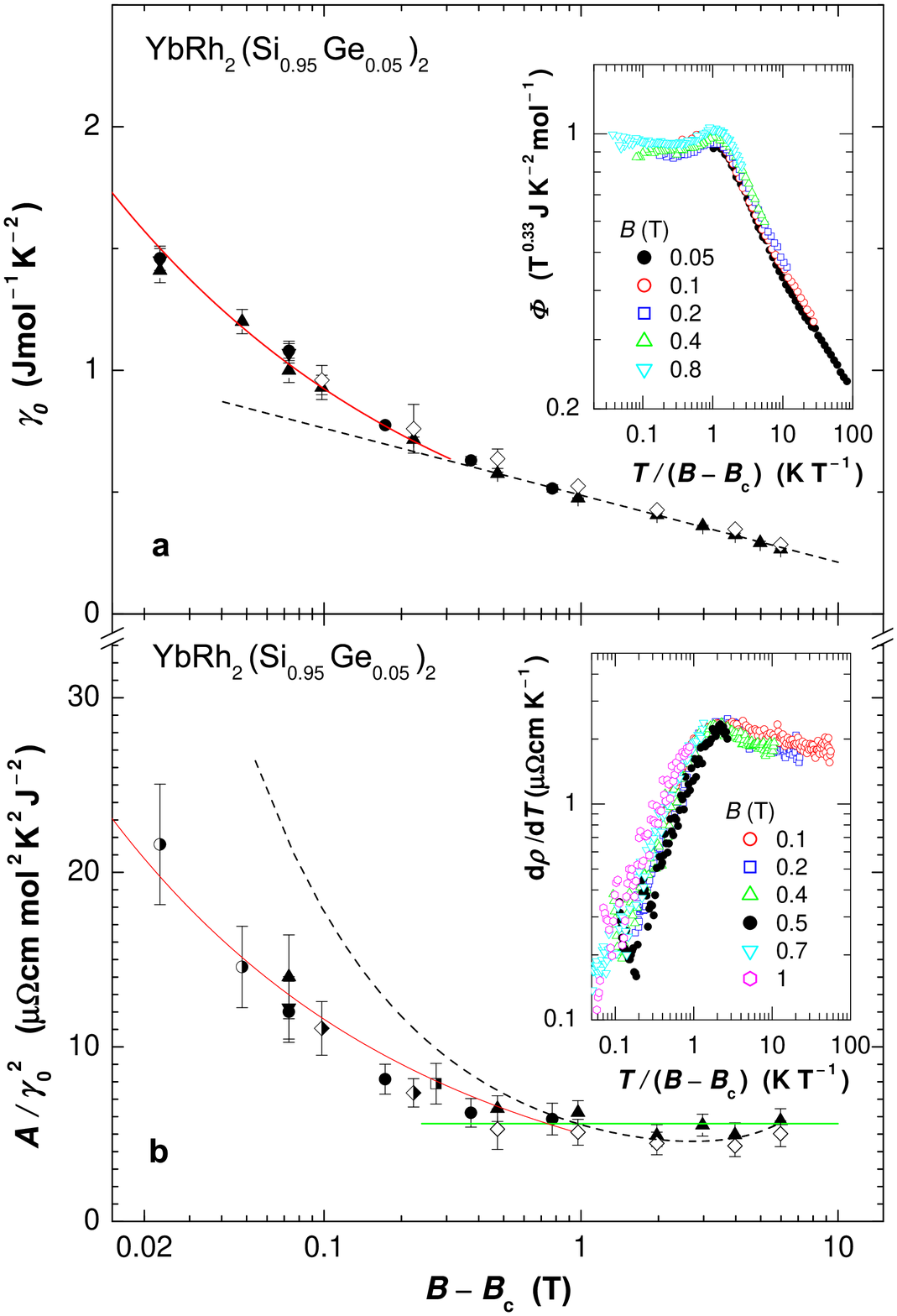}            %commentaar:Name of Figure changed!!!%
 \end{center}
\vspace*{2mm} \caption{Field dependences of the Sommerfeld
coefficient, $\gamma_{o}$, of the electronic specific heat ({\bf
a}) and of the ratio of the $A$--coefficient in the $T^{2}$ term
of the electrical resistivity and $\gamma_{o}^{2}$ ({\bf b}) for
YbRh$_{2}$(Si$_{0.95}$Si$_{0.05}$)$_{2}$. Note that $\gamma_{o}$
and $A$ are proportional to the effective quasiparticle mass and
the effective quasiparticle--quasiparticle scattering cross
section, respectively. The magnetic field was applied
perpendicular to the $c$--axis, and the applied field values are
corrected, on the abscissae, by the value of the critical field,
$B_{\rm c}=0.027$~T. $\gamma_{o}$--values in ({\bf a}) were
obtained from two different samples: Three independent
measurements on sample \#1 are displayed by closed symbols
(circles, up and down triangles). The open symbols (diamonds)
show the results of measurements on sample \#2. As $B \rightarrow
B_{\rm c}$, $\gamma_{o}$ diverges $\propto (B-B_{\rm c})^{-0.33}$
(red line), i.\,e., much stronger than logarithmically (black
dashed line). The symbols used in the semi--log plot
$K=A/\gamma_{o}^{2}$ vs $\left(B-B_{\rm c}\right)$ of ({\bf b})
correspond to values for the electronic specific heat coefficient
shown in ({\bf a}). Half filled circles (squares) display data
for which the $A$-coefficient of the electrical resistivity was
determined by extrapolating (interpolating) $A(B-B_{\rm c})$ with
respect to $(B-B_{\rm c})$. The half filled diamond represents a
point for which the $\gamma_{o}$ value was obtained by
interpolation. The black dashed line indicates $K_{{\rm SDW}}
\propto \left[(B-B_{\rm c})\ln^{2}(B-B_{c})\right]^{-1}$ for the
2D SDW scenario~\protect\cite{paul}. This is at strong variance
from the (at $B-B_{\rm c} < 0.3$~T) experimentally observed
$K\propto \left(B-B_{\rm c}\right)^{-1/3}$, arising from the
stronger than logarithmic increase of $\gamma_{o}$ upon cooling
(red line in {\bf a}). For $\left(B-B_{\rm c}\right)> 0.3$~T, $K$
becomes field independent within the error bars at a constant
value of 5.4~$\mu\Omega$cmmol$^{2}$K$^{2}$J$^{-2}$ (green
horizontal line in {\bf b}). A similar high field behavior has
been reported previously \protect\cite{PRL2002Gegenwart} on pure
YbRh$_{2}$Si$_{2}$. Noteworthy, an almost identical value for $K$
was found. Insets show scaling behavior of the low--$T$
electronic specific heat where, according to equation (1), the
ordinate is displaying $\Phi(B,T)=\left(B-B_{\rm
c}\right)^{0.33}C_{\rm {el}}/T$, {\bf a}, as well as of the
temperature derivative of the electrical resistivity,
d$\rho$/d$T$, {\bf b}, as a function of $T/(B-B_{\rm c})$.}
\label{fig:JCNatfiguur2}
\end{figure}

We now discuss the field dependence of the electronic specific
heat in YbRh$_{2}$(Si$_{0.95}$Ge$_{0.05}$)$_{2}$ in more detail.
In these measurements, magnetic fields were applied perpendicular
to the crystallographic $c$-axis, within the easy magnetic plane
(Fig.~\ref{fig:specHeatacSuscep}b). At fields above 0.1~T, $C_{\rm
el}/T$ is weakly temperature independent, as expected in a
LFL~\cite{JETP1957Landau}. A weak maximum is observed in $C_{\rm           %commentaar: Landau Fermi Liquid into LFL%
el}(T)/T$ at a characteristic temperature $T_{o} (B)$ which grows
linearly with the field (inset of Fig.~\ref{fig:JCNatfiguur2}a),
indicating that entropy is transferred from the low-temperature
upturn to higher temperatures by the application of a field $B
\geq B_{\rm c} = 0.027$~T. As the field is lowered the
temperature window over which $C_{\rm el} (T,B)/T=\gamma_{o} (B)$
is constant shrinks towards zero and the zero-temperature
$\gamma_{o} (B)$ diverges (Fig.~\ref{fig:JCNatfiguur2}a). For
example, in a field of 0.05~T a constant value $\gamma_{o}(B)
\approx 1.54(7)$~Jmol$^{-1}$K$^{-2}$ only develops below
$40$~mK.  These results indicate the formation of a field-induced
LFL state at a characteristic scale $T \ltappr T_{o} (B)$. As the
window of LFL behavior is reduced towards zero, an ever
increasing component of the zero-field upturn in the specific
heat coefficient is
revealed in the temperature dependence. This confirms that the   %commentaar:themajor separated%
major part of the upturn in the specific heat coefficient
observed in zero field is electronic in character, and must be
associated with the intrinsic specific heat at the QCP.

This conclusion is also supported by the electrical resistivity
data which reveal a field-dependent cross-over from a $T$-linear
resistivity at high temperatures, to quadratic behavior $\Delta
\rho = A(B)T^2$ at sufficiently low temperatures. Most
importantly, the data show that at low fields and temperatures,
the same scale $T_{o} (b)\propto b$ (where $b= B-B_{\rm c}$ is the       %commentaar: subscript "c" in roman style%
deviation from the critical field) governs the cross-over from LFL       %commentaar: Fermi became LFL%
to NFL behavior in {\sl both} the thermodynamics {\sl and } the          %commentaar: non-Fermi liquid became NFL%
resistivity. This can be quantitatively demonstrated by noting
that the finite field transport and specific heat data collapse
into a single set of scaling relations (see
Fig.~\ref{fig:JCNatfiguur2} inserts),

\begin{equation}\label{nn3}
\frac{C_{V}}{T}=\frac{1}{ b^{1/3}}\Phi \left(\frac{T}{T_{o} (b)}       %commentaar:subscript changed in Capital letter for consistency%
\right),\qquad \frac{d\rho }{dT}=F\left(\frac{T}{T_{o} (b)}
\right),
\end{equation}
where $\Phi (x)\sim({max}(x,1))^{-1/3}$ and $F (x)\sim x/{{max
(x,1)}}$. The NFL physics  is described by the $x\rightarrow           %commentaar: non-Fermi liquid became NFL%
\infty$ ($T>>T_{o} (b)$) behavior of these equations, where
$d\rho/dT$ is constant and $C_{V}/T\propto T^{-1/3}$. By
contrast, the field-tuned LFL is described by the $x\rightarrow 0$       %commentaar: Fermi liquid became LFL%
limit of these equations. Were there any residual pockets of LFL
behavior that were left unaffected by the QCP, we would expect a
residual quadratic component in the resistivity, and the data
would not collapse in the observed fashion. We are thus led to
believe that the break up of the LFL involves the entire Fermi
surface.

From the second scaling  relation in (\ref{nn3}),  we see that
the $A$-coefficient of the $T^{2}$ term to the resistivity              %commentaar: T^2 term added to text%
diverges roughly as $1/b$, a result that is consistent with
earlier measurements on pure
YbRh$_{2}$Si$_{2}$ carried out further away from the QCP             %commentaar: YbRh2Si2 written in the usual way%~\cite{PRL2002Gegenwart}. Over the same range, the specific heat
coefficient $\gamma_{o}(b)$ grows as
$b^{-1/3}$(Fig.~\ref{fig:specHeatacSuscep}a). Notice that the
field dependence at absolute zero temperature can be interchanged
with the temperature dependence at $B=B_{\rm c}$, but only in the
upturn region. At high magnetic field deviations from the QCP,
earlier measurements showed~\cite{PRL2002Gegenwart} that the
Kadowaki Woods ratio~\cite{SSCom1986Kadowaki} $K=A/\gamma_{o}^{2}
$ is approximately constant. Closer to the QCP,  where the
scaling behavior is observed, $K = A/\gamma_{o}^{2} \approx
b^{-1/3}$ is found to contain a weak field dependence
(Fig.~\ref{fig:JCNatfiguur2}b).                                       %commentaar: Fig 3 as reference added%

We now turn to discuss the broader implications of our
measurements. The observed divergence of both the $A$-coefficient
and the coefficient $\gamma_{o}$ of the $T$-linear specific heat
certainly rule out a 3D SDW scenario, which predicts that both              %commentaar: "that" added%
quantities will remain finite at sufficiently low temperature in
the approach to a zero-field QCP ($B \rightarrow B_{\rm c}
\rightarrow 0$), but it can be used to obtain still more insight
into the underlying scattering mechanisms between the
quasiparticles. In a 2D SDW scenario, the scattering amplitude
between two heavy electrons is severely momentum dependent. When
used to compute the transport relaxation rate, the SDW scenario
leads to the result $A\propto 1/\kappa ^{2}$, with $\kappa$ the
inverse correlation length~\cite{paul}. The observed divergence in
$A(b)$ would require $\kappa^{2}\propto b$. However, the
fluctuations of the soft 2D spin fluctuations only produce a weak
logarithmic renormalization in the heavy electron density of
states, measured by the specific heat coefficient, $\gamma_{o}
\propto \ln \left(1/\kappa\right)$.  Thus the 2D SDW scenario
predicts a weak divergence in the in--$T$ linear specific heat,
but a strongly field dependent enhancement of the Kadowaki Woods
ratio in the approach to a QCP ($b\rightarrow 0$), given by

\begin{equation}
 \gamma_{SDW}\propto \ln (1/b) \qquad {\mbox { and }} \qquad K_{SDW}\propto \frac{1}{b\ln ^{2} (b) }.
\end{equation}
The strong violation of these predictions by our data, presented
in Fig.~\ref{fig:JCNatfiguur2}a and \ref{fig:JCNatfiguur2}b, rules      %commentaar:rule changed in rules%
out 2D spin fluctuations as the driving force behind the
thermodynamics and the dominant source of scattering near the
heavy electron QCP.

Taking a more general view, scaling behavior of the transport
scattering rate tells us that the only scale entering into the
density of states and the scattering amplitude is the single
scale $T_{o}\propto b$ of the heavy electron fluid. A truly
field--independent Kadowaki Woods ratio would indicate that the
quasiparticle scattering amplitude has the form

\begin{equation}
A^{*}= T_{F}^{*}{\cal  F}[\{k_{in}\}\rightarrow \{k_{out} \}].
\end{equation}
The weak field dependence of the Kadowaki Woods ratio over a wide
range of fields implies that the characteristic length scale of
the most singular scattering amplitudes renormalizes more slowly
in the approach to the QCP than expected in a SDW scenario.                  %commentaar: "That" changed in "than"%

Our data also provide some insight into the thermodynamics in the vicinity of the QCP. By integrating the scaling
form (\ref{nn3}) for the specific heat over temperature, the entropy $S (T) = \int_{o}^{T} dT' (C_{V}/T')$ in the
vicinity of the QCP can be described by the form

\begin{equation}
S (T,B)= b^{1-\eta }{\cal  S} \left(\frac{T}{T_{o} (b)} \right)
\end{equation}
where $\eta =1/3$. The appearance of a field--dependent
pre--factor in this equation forces the entropy to vanish at the
QCP, as required by the third law of thermodynamics. The exponent
in the pre-factor also determines the effective Fermi temperature
$T_{F}^{*} (b)\propto \gamma (b)^{-1}\propto T_{o} (b)^{\eta }$.
Thus the requirement that the entropy vanishes at the QCP ($\eta
<1$) prevents a direct proportionality between the Fermi
temperature of the heavy LFL and the scale $T_{o} (b)$ governing         %commentaar: fermi liquid changed by LFL%
the cross-over to NFL behavior. It follows that the Fermi
temperature and cut-off temperature $T_o(b)$ must obey a
relationship of the form

\begin{equation}
T_F^*(b) = T_{\Lambda} \left(\frac{T_o(b)}{T_{\Lambda}}\right)^{\eta }
\end{equation}
where $T_{\Lambda}$ is an upper cut-off that we might identify
with the single ion Kondo temperature of the Yb$^{3+}$               %commentaar:Ytterbium replaced by Yb3+%
ions($\approx 25K$). Such a power law renormalization of the
characteristic energy scale would be expected  in the presence of
locally critical fluctuations that extend down from $T_{\Lambda}$
to the infra-red cut--off provided, in this case, by the magnetic
field\cite{noz2}.

In this respect, our results support the conclusions recently drawn from earlier measurements on the quantum
critical material $CeCu_{6-x}Au_{x}$ ($x=0.1$) \cite{NAT2000Schroeder}, and used in a recently proposed theory
for quantum criticality by Si {\em et al.} \cite{NAT2001Si}, suggesting that the most critical scattering is
neither three, two or even one dimensional, but local-- as if the most critical fluctuations in the underlying
quantum phase transition are fundamentally ``zero dimensional'' in character.

One of our most striking observations is that below $T \approx
0.3$~K where $\chi(T)$ follows a Curie-Weiss law, the electronic
specific heat coefficient $C_{\rm {el}}(T) /T$ for both samples
starts to deviate towards larger values, separating away from the
$-\log T$ dependence that is valid\cite{PRL2000Trovarelli} up to
10~K (Fig.~\ref{fig:specHeatacSuscep}a). This ``upturn''
continues in the $x=0.05$ sample down to approximately 20~mK, if
the critical field of 0.027~T ($B \perp c$) is applied. We
ascribe this intrinsically electronic feature to the critical
fluctuations associated with the zero-field quantum phase
transition that exists at a slightly larger Ge concentration. The
unique temperature dependence of $C_{\rm {el}}(T)/T$ for
$T<0.3$~K is {\em disparate} from the linear temperature
dependence of the electrical resistivity which holds all the way
from $\geq 10$~K to $T \approx 10$~mK. Since the former
(thermodynamic) quantity probes the dominating local 4{\em f\/}
(``spin'') part of the composite quasiparticles , while the
latter (transport) quantity is sensitive to the itinerant
conduction-electron (``charge'') part, one may view the observed
disparity as a direct manifestation of the {\em break up} of the
composite fermion in the approach to the QCP.

\newpage

{\bf Acknowledgements}

We would like to acknowledge discussions with J. Ferstl,
C. Langhammer, S. Mederle, N. Oeschler, I. Zerec, G.  Sparn,
O. Stockert, M. Abd-Elmeguid, J. Hopkinson, A. I. Larkin, and
I. Paul. Work at Dresden is partially supported by the Fonds der
Chemischen Industrie and by the FERLIN project of the European Science
Foundation. P.  Coleman is supported by the National Science
Foundation. Y.~Tokiwa is a Young Scientist Research Fellow supported
by the Japan Society for the Promotion of Science.

\newpage

%################################################################

%

%           FIGURE Exponentplot for x=0 and x=0.05

%

%

%################################################################

%################################################################

%%

%%          FIGURE C/T vs log T  and Xac for x=0 and x=0.05

%%                 (a) Xac

%%                 (b) C/T vs log T; inset crystal structure

%%

%%################################################################

\newpage

%################################################################

%

%           FIGURE A-coefficient vs B  and  gamma vs B  for x=0.05

%

%

%################################################################

\newpage











\begin{thebibliography}{99}



\bibitem{PRB1976Hertz} Hertz, J.A., Quantum critical phenomena. {\em Phys. Rev. B} {\bf 14,} 1165-1184 (1976).

\bibitem{PRB1993Millis} Millis, A.J., Effect of a nonzero temperature on quantum critical points in itinerant fermion systems.
{\em Phys. Rev. B} {\bf 48,} 7183-7196 (1993).

\bibitem{PR1994Continentino} Continentino, M.A., Quantum scaling in many-body systems.
{\em Phys. Rep.} {\bf 239,} 179-213 (1994).

\bibitem{NAT2001Si} Si, Q., Rabello, S., Ingersent, K. \& Smith, J.L., Locally critical quantum phase transitions in strongly correlated metals.
{\em Nature} {\bf 413,} 804-808 (2001).

\bibitem{qcp}  Coleman, P. \& P{\'e}pin, C., What is the fate of the heavy electron at a quantum critical point?
{\em Physica B} {\bf 312,} 383-389 (2002).



\bibitem{PRL2002Gegenwart} Gegenwart, P. {\em et al.,} Magnetic-field induced quantum critical point in YbRh$_{2}$Si$_{2}$.
{\em Phys.~Rev.~Lett.} {\bf 89,} 056402 (2002).

\bibitem{greg}Heuser, K. {\em et al.,} Inducement of non-Fermi-liquid behavior with a magnetic field.
{\em Phys Rev. B} {\bf 57,} R4198-201, (1998).


\bibitem{hilbert}  Stockert, O.  {\em et al.,}
Pressure versus magnetic-field tuning of a magnetic quantum phase transition. {\em Physica B} {\bf 312-313},
458-460 (2002).

\bibitem{PRL2000Trovarelli} Trovarelli, O. {\em et al.,}
YbRh$_{2}$Si$_{2}$: Pronounced non-Fermi-liquid effects above a low-lying magnetic phase transition.
Phys.~Rev.~Lett. {\bf 85,} 626-629 (2000).

\bibitem{JPCM2002Mederle} Mederle, S. {\em et al.,} Unconventional metallic state in YbRh$_{2}$(Si$_{1-x}$Ge$_{x}$)$_{2}$ - a high pressure study.
{\em J.~Phys.:~Condens.~Matter} {\bf 14,} 10731-10736 (2002).

\bibitem{Abd-Elmeguid1999} Plessel, J., Hochdruckuntersuchungen an Yb-Kondo-Gitter-Systemen in der N{\"a}he einer magnetischen Instabilit{\"a}t.
 {\em Dissertation, Universit{\"a}t zu K{\"o}ln} (2001).

\bibitem{Francois1985} Francois, M., Venturini, G., March{\^e}ch{\'e}, J.F., Malaman, B. \& Roques, B.,
De Nouvelles s{\'e}ries de germaniures, isotopes de U$_{4}$Re$_{7}$Si$_{6}$, ThCr$_{2}$Si$_{2}$ et
CaBe$_{2}$Ge$_{2}$, dans les syst{\`e}mes ternaires R--T--Ge o{\`u} R est un {\'e}l{\'e}ment des terres rares et
T $\equiv$ Ru, Os, Rh, Ir: supraconductivit{\'e} de LaIr$_{2}$Ge$_{2}$. {\em J.~Less Common Metals} {\bf 113}
231-237 (1985).

\bibitem{SSCom1986Kadowaki} Kadowaki, K. \& Woods, S.B., Universal relationship of the resistivity and specific heat in heavy-fermion compounds.
{\em Solid~State~Commun.} {\bf 58,} 507-509 (1986).


\bibitem{APPB2003Gegenwart} Gegenwart, P. {\em et al.,} Divergence of the heavy quasiparticle
mass at the antiferromagnetic quantum critical point in YbRh$_{2}$Si$_{2}$. {\em Acta Phys. Pol. B} {\bf 34,}
323-334 (2003).

\bibitem{JETP1957Landau} Landau, L.D., The theory of a Fermi liquid. {\em Sov. Phys. JETP} {\bf 3,} 920-925
(1957).

\bibitem{paul}  Paul, I. \& Kotliar, G., Thermoelectric behavior near the magnetic quantum critical point. {\em Phys.~Rev.~B} {\bf 64,} 184414 (2001).

\bibitem{noz2} Giamarchi, T. , Varma, C.M., Ruckenstein, A.E. \& Nozi{\`e}res, P., Singular low energy
properties of an impurity model with finite range interactions. {\em Phys. Rev. Lett.} {\bf 70}, 3967- 3970
(1993).

\bibitem{NAT2000Schroeder} Schr{\"o}der, A. {\em et al.,} Onset of antiferromagnetism in heavy-fermion metals.                             {\em Nature} {\bf 407,} 351-355 (2000).


\bibitem{PMS1977Carter} Carter, G.C.  {\em et al.,} in Metallic shifts in NMR.
{\em Progress in Materials Science,} {\bf Vol. 20, Part I,} ed.
Chalmers, B., Christian, J.W. \& Massalski, T.B. (Pergamon Press,
Oxford 1977).




%\bibitem{ACTA2001Custers} J.~Custers {\em et al.,} Low-temperature magnetic and transport properties of the clean NFL system YbRh$_{2}$(Si$_{1-x}$Ge$_{x}$)$_{2}$.
%{\em Act.~Phys.~Pol.~B} {\bf 32,} 3221-3217 (2001).

%\bibitem{PRL1998Schroeder} Schr{\"o}der, A., Aeppli, G., Bucher, E., Ramazashvili, R. \& Coleman, P.,
%Scaling of magnetic fluctuations near a quantum phase transition. {\em Phys.~Rev.~Lett.} {\bf 80} 5623-5626
%(1998).

%%\bibitem{CondMat2002Ishida} Ishida, K. {\em et al.,}

%%                                Spin dynamics in a structurally ordered non-Fermi liquid compound: YbRh$_{2}$Si$_{2}$.

%%                                {\em cond-mat}/0207464v1 (18 July 2002).


%\bibitem{JPSJ1996Kambe} Kambe, S. {\em et al.,} Application of the SCR spin fluctuation theory for the magnetic instability in heavy fermion system Ce$_{1-x}$La$_{x}$Ru$_{2}$Si$_{2}$.
%{\em J. Phys. Soc. Jpn.} {\bf 65,} 3294-3300 (1996).

%%\bibitem{PRL1997Rosch} Rosch, A., Schr{\"o}der, A., Stockert, O., \& L{\"o}hneysen, H.v., Mechanism for the non-Fermi-liquid behavior in CeCu$_{6-x}$Au$_{x}$.
%%{\em Phys.~Rev.~Lett.} {\bf 79,} 159-162 (1997).

%%\bibitem{NATURE98Mathur} Mathur, N.D. {\em et al.,} Magnetically mediated superconductivity in heavy fermion compounds.
%%{\em Nature} {\bf 394}, 39-43 (1998).

%%\bibitem{PRL1994Loehneysen} L{\"o}hneysen, H.v. {\em et al.,} Non-Fermi-liquid behavior in a heavy-fermion alloy at a magnetic instability.
%%{\em Phys.~Rev.~Lett.} {\bf 72,} 3262-3265 (1994).

%%\bibitem{JPCM2000Grosche}  Grosche, F.M. {\em et al.,} Anomalous low temperature states in CeNi$_2$Ge$_2$ and CePd$_2$Si$_2$.
%%{\em J. Phys.: Condens. Matter} {\bf 12}, L533-L540 (2000).


%%\bibitem{KuechlerUP}  K{\"u}chler, R., Weickert, F. \& Gegenwart, P., {\em unpublished results}.


%\bibitem{PB2003Steglich} F. Steglich {\em et al.,}

%                             Recent trends in heavy-fermion physics.

%                             {\em Physica B} (forthcoming).



%\bibitem{SakakibaraJJAP1994} T. Sakakibara {\em et al.,}

%                                 Faraday force magnetometer for high-sensitivity magnetization measurements at very low temperatures and high fields.

%                                 {\em Jpn.~J.~Appl.~Phys.} {\bf 33,} 5067-5072 (1994).

%\bibitem{KuechlerUP}  R. K{\"u}chler and F. Weickert,

%                          {\em unpublished results}.



\end{thebibliography}
\end{document}